\documentstyle[12pt]{article}
\def\nline{\,\nabla\kern -0.7em\raise0.2ex\hbox{/}\,\,}
\def\yline{\,y\kern -0.47em /}
\def\aline{\,a\kern -0.49em /}
\def\parline{\,\partial\kern -0.55em /\,\,}
\def\vphhh{{\vphantom{5pt}}}

\def\NPB#1(#2)#3{{\it Nucl. Phys.} {\bf B#1} (#2) #3}
\def\PRD#1(#2)#3{{\it Phys. Rev.} {\bf D#1} (#2) #3}
\def\PLB#1(#2)#3{{\it Phys. Lett.} {\bf B#1} (#2) #3}

\def\RMP#1(#2)#3{{\it Rev. Mod. Phys.} {\bf #1} (#2) #3}
\def\MPLA#1(#2)#3{{\it Mod. Phys. Lett.} {\bf A#1} (#2) #3}
\def\CQG#1(#2)#3{{\it Class. Quantum Grav.} {\bf #1} (#2) #3}
\def\AP#1(#2)#3{{\it Ann. Phys.} {\bf #1} (#2) #3}
\def\SJNP#1(#2)#3{{\it Sov. J. Nucl. Phys.} {\bf #1} (#2) #3}

\begin{document}

{}~
\vspace{1cm}
\begin{center}

{\Large\bf
Fermionic fields in
the d-dimensional anti-de Sitter

\medskip
spacetime}

\vspace{2cm}
R.R. Metsaev

\vspace{1cm}
{\it Department of Theoretical Physics, P.N. Lebedev Physical
Institute, Leninsky prospect 53, 117924, Moscow, Russia}

\vspace{3cm}
{\bf Abstract}
\end{center}

\noindent Arbitrary spin free massless fermionic fields
corresponding to mixed symmetry representations of the
\hbox{$SO(d-1)$} compact group and propagating in even
$d$-dimensional anti-de Sitter spacetime are investigated.
Free wave equations of motion, subsidiary conditions and the
corresponding gauge transformations for such fields are
proposed.  The lowest eigenvalues of the energy operator for the
massless fields and the gauge parameter fields are
derived. The results are formulated in $SO(d-1,2)$ covariant
form as well as in terms of intrinsic coordinates.

\vspace{3cm}

\centerline{hep-th/9802097, published in {\it Phys. Lett.} B419 (1998) 49}

\newpage

In view of the aesthetic features of anti-de Sitter field
theory a interest in this theory was periodically renewed
(see \cite{F1}-\cite{Nic} and references therein). One of
the interesting directions of this theory is the higher-spin
massless field theory. At present there is reason for revival of
interest in higher-spin massless fields.  Recently it was
discovered \cite{Vas1} that the consistent equations of motion of
interacting higher-spin massless fields in four dimensional
($d=4$) anti-de Sitter spacetime could be set up with the help
of higher-spin superalgebras (for review see \cite{Vas4}).
These equations for the case of higher spacetime dimensions
$d>4$ have been generalized in \cite{Vas2}.  Because these
equations are formulated in terms of wavefunctions which depend
on usual spacetime coordinates and certain twistor variables it
is not clear immediately what kind of fields they describe.  In
\cite{Vas1}, it was established that in four dimensions
($d=4$) they describe unitary dynamics of massless fields of all
spins in anti-de Sitter spacetime.

The long-term motivation of our investigation
is to provide an answer to question:
do the higher dimensional theories of Ref.\cite{Vas2} describe
higher-spin massless fields?
The first step in such an investigation is a description of free
equations of motion for arbitrary spin massless fields. In
contrast to completeness of description for $d=4$
(\cite{F1},\cite{Ish1},\cite{Nic},\cite{FH}-\cite{BrFr}), not
much is known on the higher-spin fermionic massless fields for
$d>4$ even at the level of free fields, unless
considerations are restricted to totally (anti)symmetric
representations (\cite{Vas3}-\cite{Metsit4}).
The present paper is a sequel to our paper \cite{Metsit2} where
description of bosonic massless fields of all spins in anti-de
Sitter spacetime of arbitrary dimensions was developed.  The
case of half-integral spins, presented here is a necessary step
in our study of massless fields of all spins for arbitrary
$d$.  With higher-spin theories in mind we also hope that our
results may be of wider interest, especially to anti-de Sitter
supergravity theories.

Let us first formulate the main problems we solve in this
letter. A positive-energy lowest weight
irreducible representation of the $SO(d-1,2)$ group denoted
as $D(E_0, {\bf m})$, is defined by  $E_0$,
the lowest eigenvalue of
the energy operator, and by ${\bf m}=(m_1^\vphhh,\ldots
m_\nu^\vphhh)$, $\nu\equiv\frac{d-2}{2}$, which is the weight of
the unitary representation of the $SO(d-1)$ group.
The $m_i$  are half integers for fermions. For the
case of $SO(3,2)$ group it has been discovered \cite{FH} that
fermionic massless fields propagating in four dimensional
anti-de Sitter spacetime are associated with $D(s+1,s)$ i.e.
spin-$s$ anti-de Sitter massless particle takes lowest
value of energy equal to $E_0=s+1$.  In \cite{Metsit4} for the
case of arbitrary $d$  it was found that $E_0=s+d-3$ for
totally symmetric massless representation
${\bf m}_{sym}=(s,1/2,\ldots ,1/2)$, $s\geq 3/2$,  and
$E_0=d-s^\prime-1/2$
for the totally antisymmetric one ${\bf m}_{as}=(3/2,\ldots
,3/2,1/2\ldots 1/2)$ (where the 3/2 occurs $s^\prime\le \nu $
times in this sequence).  The cases ${\bf m}_{sym}$ and ${\bf
m}_{as}$ are very special:  there are many representations
corresponding to arbitrary ${\bf m}$ (mixed symmetry
representations).  In this paper we construct free equations of
motion and gauge transformations for massless fields with
arbitrary ${\bf m}$ and determine the $E_0$.  Here we restrict
ourselves to the case of even spacetime dimension $d$.

Let us describe our conventions and
notation.  We describe the anti-de Sitter spacetime as a
hyperboloid
\begin{equation}\label{hyperbol}
\eta_{AB}^\vphhh y^A y^B=-1\,,
\qquad
\eta_{AB}=(-,+,\ldots,+,-)\,,
\qquad
A,B=0,1,\ldots,d-1,d+1\,,
\end{equation}
in $d+1$- dimensional pseudo-Euclidean space with metric tensor
$\eta_{AB}$. The indices $A,B$ are raised and lowered by $\eta^{AB}$
and $\eta_{AB}$ respectively.  In what follows to simplify our
expressions we will drop the metric tensor
$\eta_{AB}^\vphhh$ in scalar products.  The generators
$J^{AB}$ of the $SO(d-1,2)$ group satisfy the commutation
relations

$$
[J^{AB},J^{CD}]=\eta^{BC}J^{AD}+3 \hbox{ terms}\,.
$$
As is usual, we split $J^{AB}$ into an orbital part
$L^{AB}$ and a spin part $M^{AB}$:
$J^{AB}=L^{AB}+M^{AB}$. The realization
of $L^{AB}$ in terms of differential operators defined
on the hyperboloid (\ref{hyperbol}) is:
$$
L^{AB}=y^A \nabla^B-y^B \nabla^A\,,
\qquad
\nabla^A\equiv\theta^{AB}\frac{\partial}{\partial y^B}\,,\qquad
\theta^{AB}\equiv \eta^{AB}+y^Ay^B\,,
$$
The tangent derivative $\nabla^A$ has the properties
$y^A\nabla^A=0$, $\nabla^A y^A=d$
and satisfies the commutation relations

$$
[\nabla^A,y^B]=\theta^{AB}\,,
\qquad
[\nabla^A,\nabla^B]=- L^{AB}\,.
$$
A form for $M^{AB}$ depends on the
realization of the representations. We will use the tensor
realization of representations. As the carriers for
$D(E_0,{\bf m})$ we use tensor-spinor field

$$
\Psi^{A({\bf m})}=\Psi_{\phantom{1}}^{A^1_1,\ldots,
A^{m^\prime_1}_1,\ldots, A^1_\nu,\ldots,
A^{m^\prime_{_\nu}}_\nu}\,, \qquad m_i^\prime\equiv
m_i-\frac{1}{2}\,, $$
defined on the hyperboloid  (\ref{hyperbol}). The $\Psi$ has one
spinor index which we do not show explicitly. By definition,
$\Psi^{A({\bf m})}$ is a tensor-spinor field whose $d+1$
spacetime indices $A$ have the structure of the Young tableaux
($YT$) corresponding to the irreps of the $SO(d-1,2)$ group
labeled by ${\bf m}^\prime=(m_1^\prime\ldots m_\nu^\prime)$. In
what follows we use the notation ${\sf YT}({\bf m^\prime})$ to
indicate such $YT$.  At the same time the ${\bf m}$ is
the weight of a representation of the $SO(d-1)$ group, and the
$m_i$ satisfy the inequalities

\begin{equation}\label{mineq1}
m_1^\vphhh \geq\ldots\geq
m_\nu^\vphhh \geq 1/2\,.
\end{equation}
In the language of $YT$, the
$m_i^\prime$ indicates the number of boxes in the $i$-th row of
${\sf YT}({\bf m}^\prime)$.  Thus given an arbitrary lowest
weight massless representation $D(E_0,{\bf m})$ we assume that a
covariant description can be formulated with the field
$\Psi^{A({\bf m})}$. To avoid cumbersome tensor expressions we
introduce $\nu$ creation and annihilation operators $a_l^A$ and
$\bar{a}_l^A$ ($l=1,\ldots \nu$, $\nu=(d-2)/2$)
which satisfy

$$
[\bar{a}_i^A,a_j^B]=
\eta^{AB}\delta_{ij}\,,\qquad
\bar{a}_i^A|0\rangle=0
$$
and construct a Fock space vector

\begin{equation}\label{fockv}
|\Psi\rangle \equiv \prod_{l=1}^\nu\prod_{i_l=1}^{m^\prime_l}
a_l^{A_l^{i_l}}
\Psi_{\phantom{1}}^{A^1_1,\ldots, A^{m^\prime_1}_1,\ldots,
A^1_\nu,\ldots, A^{m^\prime_{_\nu}}_\nu}|0\rangle\,.
\end{equation}
For a realization of this kind, $M^{AB}$ has the form

$$
M^{AB}=M^{\prime AB}+\sigma^{AB}\,,
$$
where
$$
M^{\prime AB}= \sum_{l=1}^\nu
(a_l^A\bar{a}_l^B-a_l^B\bar{a}_l^A)\,,
\qquad
\sigma^{AB}=\frac{1}{4}(\gamma^A\gamma^B-\gamma^B\gamma^A)
$$
and $\gamma^A$ are the gamma matrices:
$\{\gamma^A,\gamma^B\}=2\eta^{AB}$.
Throughout of the paper, unless otherwise specified, the indices
$i,j,l,n$ run over $1,\ldots, \nu$. For these indices
we drop the summation over repeated indices.

If the $\Psi^{A({\bf m})}$ is associated
with ${\sf YT}({\bf m}^\prime)$ then the $|\Psi\rangle$
satisfies

\begin{equation}\label{ytfcon1}
(a_{ii}^\vphhh-m_i^\prime)|\Psi\rangle=0\,,\quad
a_{ij}^-|\Psi\rangle=0\,,\quad
\varepsilon^{ij}_\vphhh a_{ij}^\vphhh|\Psi\rangle=0\,,
\end{equation}
where in (\ref{ytfcon1}) and below we use the notation
$$
a_{ij}^\vphhh\equiv  a_i^A \bar{a}_j^A\,,
\qquad
a_{ij}^-\equiv \bar{a}_i^A \bar{a}_j^A\,,
\qquad
a_{ij}^+\equiv a_i^A a_j^A\,,
$$
and $\varepsilon^{ij}=1(0)$ for $i<j(i\geq j)$.
The 1st equation in (\ref{ytfcon1}) tells us that
$a_i$ occurs $m_i^\prime$ times on the right hand side of
eq.(\ref{fockv}). Tracelessness of $\Psi^{A({\bf m})}$
is reflected in the 2nd equation in (\ref{ytfcon1}).
The 3rd equation in (\ref{ytfcon1})
is a consequence of the following Young's symmetrization rule we
use:  (i) first we perform alternating
with respect to indices in
all columns, (ii) then we perform symmetrization with respect
to indices in all rows  (for more explanations see
\cite{Zhel})

Being the carrier for the $D(E_0, {\bf m})$, the field
$|\Psi\rangle$ satisfies the equation

\begin{equation}\label{eqmot1}
(Q-\langle Q\rangle )|\Psi\rangle=0\,,
\qquad
Q\equiv\frac{1}{2} J^{AB}J^{AB}
\end{equation}
and allows the following subsidiary covariant constraints

\begin{eqnarray}
\label{diver}
\bar{\nabla}_n|\Psi\rangle&=&0\,,\quad
(\hbox{divergenelessness})
\\
\label{ytrans}
\bar{y}_n|\Psi\rangle&=&0\,,\quad (\hbox{transversality})
\\
\label{gtrans}
\bar{\gamma}_n|\Psi\rangle&=&0\,,
\end{eqnarray}
$n=1,\ldots.\nu$.
In (\ref{eqmot1}) $Q$ is the second
order Casimir operator of the $so(d-1,2)$ algebra
while $\langle Q\rangle$ is its eigenvalue for
$D(E_0, {\bf m})$
$$
\langle Q\rangle =-E_0(E_0+1-d)
-\sum_{l=1}^\nu m_l^\vphhh (m_l^\vphhh -2l +d-1)\,,
$$
where $E_0$ is the lowest eigenvalue of ${\rm i}J^{d+10}$.
The $\langle Q\rangle$ can be calculated according to the well
known procedure \cite{PerPop}.
In (\ref{diver})-(\ref{gtrans}) and
below we use the notation

$$ \nabla_n\equiv {\bf a}_n^A \nabla^A\,,\qquad
\bar{\nabla}_n
\equiv \nabla^A {\bf a}_n^A\,,
\qquad
{\bf a}_n^A\equiv \theta^{AB}a_n^B\,,\qquad
\bar{{\bf a}}_n^A\equiv \theta^{AB}\bar{a}_n^B\,.
$$

$$
y_n\equiv  y^A a_n^A\,,\qquad
\bar{y}_n\equiv y^A\bar{a}_n^A\,,
\qquad
\gamma_n\equiv \gamma^A a_n^A\,,
\qquad
\bar{\gamma}_n\equiv \gamma^A \bar{a}_n^A\,.
$$
Introducing
$$
{\bf a}_{ij}\equiv  {\bf a}_i^A\bar{\bf a}_j^A\,,\quad
{\bf a}_{ij}^-\equiv \bar{\bf a}_i^A\bar{\bf a}_j^A\,,\quad
{\bf a}_{ij}^+\equiv  {\bf a}_i^A {\bf a}_j^A
$$
and using the constraints (\ref{ytrans}) and the equalities

\begin{equation}\label{oscdecom}
a_{ij}^\vphhh={\bf a}_{ij}-y_i\bar{y}_j\,,\qquad
a_{ij}^-={\bf a}_{ij}^- -\bar{y}_i\bar{y}_j,
\end{equation}
the constraints (\ref{ytfcon1}) can be rewritten in the form
which we use in what follows:

\begin{equation}\label{ytfcon2}
({\bf a}_{ii}-m_{i}^\prime)|\Psi\rangle=0\,,\quad
{\bf a}_{ij}^-|\Psi\rangle=0\,,\quad
\varepsilon^{ij}_\vphhh{\bf a}_{ij}|\Psi\rangle=0\,.
\end{equation}
For $E_0$ corresponding to the massless case a
solution of Eqs. (\ref{eqmot1})-(\ref{gtrans}) decomposes into a
physical massless representations and some additional
representations corresponding to pure gauge fields (i.e. the
situation is similar to the $d=4$ case \cite{FF}).

Now rewriting the expression for $Q$ as

$$
Q=-\nabla^2+M^{AB}L^{AB}+\frac{1}{2}M^{AB}M^{AB}\,,
\qquad
\nabla^2\equiv \nabla^A\nabla^A
$$
and taking into account
the easily derived equalities (using
eqs.(\ref{diver})-(\ref{gtrans}) and (\ref{ytfcon2}))

$$
\sigma^{AB}L^{AB}|\Psi\rangle
=\yline\nline|\Psi\rangle\,,
\qquad
M^{\prime AB}M^{\prime AB}|\Psi\rangle
=-2\sum_{l=1}^\nu
m_l^\prime(m_l^\prime-2l+d+1)|\Psi\rangle\,,
$$

$$
M^{\prime AB}L^{AB}|\Psi\rangle
=2\sum_{l=1}^\nu m_l^\prime\,|\Psi\rangle\,,
\qquad
M^{\prime AB}\sigma^{AB}|\Psi\rangle
=-\sum_{l=1}^\nu m_l^\prime\,|\Psi\rangle\,,
$$
$\nline^2=\nabla^2-\yline\nline$, where
$\nline\equiv \gamma^A\nabla^A$, $\yline\equiv \gamma^A y^A$,
$\{\nline,\yline\}=d$, we can rewrite Eq. (\ref{eqmot1}) as

\begin{equation}\label{eqmot2}
(\nline^2-(E_0+\frac{1}{2})(E_0+\frac{1}{2}-d))|\Psi\rangle=0\,.
\end{equation}
To define $E_0$ corresponding to massless fields we construct
gauge invariant equations of motion for $|\Psi\rangle$.
In order to formulate gauge transformations we use the
gauge parameters fields whose spacetime indices correspond to
the $YT$ which one can make by removing one box from the ${\sf
YT}({\bf m}^\prime)$.  Thus the most general gauge
transformations we start with are

\begin{equation}\label{gaugetr1}
\delta_{(n)}|\Psi\rangle\sim \nabla_n |\Lambda_n\rangle+
y_n |R_n\rangle+\gamma_n|S_n\rangle\,,
\end{equation}
where the gauge parameters fields
$|\Lambda_n\rangle$, $|R_n\rangle$ and $|S_n\rangle$ are
associated with ${\sf YT}({\bf m}_{(n)}^\prime)$, and $i$-th
component of the ${\bf m}_{(n)}^\prime$ is equal to
$m_{i(n)}^\prime=m_i^\prime-\delta_{in}$.  The
${\sf YT}({\bf m}_{(n)}^\prime)$ is obtained by removing one box
from $n$-th row of the ${\sf YT}({\bf m}^\prime)$.
We assume that only those $|\Lambda_n\rangle$,
$|R_n\rangle$ and $|S_n\rangle$
are non-zero whose ${\bf m}_{(n)}^\prime$ satisfy the
inequalities

\begin{equation}\label{gpwcon}
m_{1(n)}^\prime \geq\ldots \geq m_{\nu(n)}^\prime \geq 0\,.
\end{equation}
Given the ${\bf m}^\prime$, the set of those $n$ whose
${\bf m}_{(n)}^\prime$ satisfy (\ref{gpwcon}) will be referred
to as ${\sf S({\bf m}^\prime)}$, while the number of such $n$
will be referred to as $\nu\,{}^\prime$.  Thus we have
$\nu\,{}^\prime$ nontrivial gauge transformations.  We impose on
the  $|\Lambda_n\rangle$, $|R_n\rangle$ and $|S_n\rangle$ the
constraints similar to (\ref{diver})-(\ref{gtrans})

\begin{equation}\label{gpcon1}
\bar{\nabla}_i|\Lambda_n\rangle=0\,,
\qquad
\bar{y}_i|\Lambda_n\rangle=0\,,
\qquad
\bar{\gamma}_i |\Lambda_n\rangle=0\,,
\end{equation}
and constraints obtained from (\ref{gpcon1}) by replacing
$\Lambda\rightarrow R$ and $\Lambda\rightarrow S$.
Since
$|\Lambda_n\rangle$, $|R_n\rangle$ and $|S_n\rangle$ correspond
to ${\sf YT}({\bf m}_{(n)}^\prime)$, they satisfy the constraints

\begin{equation}\label{ytgpcon1}
(a_{ii}^\vphhh-m_{i(n)}^\prime)|\Lambda_n\rangle=0\,,\quad
a_{ij}^-|\Lambda_n\rangle=0\,,\quad
\varepsilon^{ij}_\vphhh a_{ij}^\vphhh|\Lambda_n\rangle=0\,,
\end{equation}
and those which are obtainable from (\ref{ytgpcon1})
by replacing $\Lambda\rightarrow R$ and
$\Lambda\rightarrow S$.
Using (\ref{oscdecom})
and (\ref{gpcon1}), the constraints (\ref{ytgpcon1}) can be
rewritten in the form similar to (\ref{ytfcon2})

\begin{equation}\label{ytgpcon2}
({\bf a}_{ii}-m_{i(n)}^\prime)|\Lambda_n\rangle=0\,,\quad
{\bf a}_{ij}^-|\Lambda_n\rangle=0\,, \quad
\varepsilon^{ij}_\vphhh{\bf a}_{ij}|\Lambda_n\rangle=0\,.
\end{equation}

Now we should find such $\delta_{(n)}|\Psi\rangle$ and $E_0$
that the constraints and the equations of
motion (\ref{diver})-(\ref{gtrans}), (\ref{ytfcon2}),
(\ref{eqmot2}) be invariant with respect to gauge
transformations.  We proceed in the following way.

(i) from the invariance requirement of (\ref{ytrans}) (i.e.
$\bar{y}_i\delta_{(n)}|\Psi\rangle=0$) and
(\ref{gaugetr1}), (\ref{gpcon1}) one gets

\begin{equation}\label{rsol}
|R_n\rangle
=-\sum_{l=1}^\nu
{\bf a}_{ln}^\vphhh|\Lambda_l\rangle
+\yline |S_n\rangle\,.
\end{equation}
By substituting (\ref{rsol}) into (\ref{gaugetr1}) we obtain

\begin{equation}\label{gaugetr2}
\delta_{(n)}|\Psi\rangle
\sim D_n|\Lambda_n\rangle
+(\gamma {\bf a}_n)|S_n\rangle\,,
\qquad
n\in {\sf S({\bf m}^\prime)}\,,
\end{equation}
where $(\gamma{\bf a}_n)\equiv\gamma^A{\bf a}_n^A$ and

\begin{equation}\label{prderiv}
D_n^\vphhh\equiv \nabla_n
+\sum_{l=1}^\nu(-y_l^\vphhh {\bf a}_{nl}
+{\bf a}_{nl}^+\bar{y}_l^\vphhh)\,.
\end{equation}
In (\ref{prderiv}) we add by hand the 2nd term in the sum. Due to
(\ref{gpcon1}) the gauge transformations (\ref{gaugetr2}) are
unaffected by this term. The $D_n$ introduced in such a way
allows the representation $D_n=a_n^Ay^BJ^{\prime AB}$, where
$J^{\prime AB}\equiv L^{AB}+M^{\prime AB}$, which is very
convenient in practical calculations.  Some useful commutation
relations are

\begin{eqnarray*}
&&{}[{\bf a}_{ij},D_n]=\delta_{jn} D_i\,,
\qquad\qquad\,\,\,\,\,
{}[{\bf a}_{ij}^-,D_n]
=\delta_{in}\bar{D}_j+\delta_{jn}\bar{D}_i\,,
\\
&&{}[{\bf a}_{ij},{\bf a}_{nl}]
=\delta_{jn}{\bf a}_{il}
-\delta_{il}{\bf a}_{nj}\,,\,
\quad
{}[{\bf a}_{ij}^-,{\bf a}_{nl}]
=\delta_{jn}{\bf a}_{il}^-
+\delta_{in}{\bf a}_{jl}^-\,,
\end{eqnarray*}
where
$\bar{D}_n\equiv J^{\prime AB}\bar{a}_n^Ay^B$. The usage
of $D_i$, $\bar{D}_i$, ${\bf a}_{ij}^\pm$ and
${\bf a}_{ij}$ is advantageous since
all of them commute with $y_n$ and $\bar{y}_n$.

(ii) using (\ref{gpcon1}), (\ref{ytgpcon2}) one can verify
that the 1st and 2nd constraints from (\ref{ytfcon2})
are invariant with respect to (\ref{gaugetr2}) i.e.  the
equations $({\bf a}_{ii}-m_i^\prime)\delta_{(n)}|\Psi\rangle=0$
and ${\bf a}_{ij}^-\delta_{(n)}|\Psi\rangle=0$ are fulfilled.
The invariance requirement of 3rd constraint from
(\ref{ytfcon2})
(i.e.
$\varepsilon^{ij}{\bf a}_{ij}\delta_{(n)}|\Psi\rangle=0$)
is the most difficult point.
Fortunately, this part of analysis is similar to that of the
bosonic case (\cite{Metsit2}). Our result for $\delta_{(n)}$ is:

\begin{equation}\label{gaugetr3}
\delta_{(n)}|\Psi\rangle
={\cal D}_n|\Lambda_n\rangle
+{\cal G}_n|S_n\rangle\,,
\end{equation}
$$
{\cal D}_n\equiv\sum_{l=1}^\nu P_{nl}D_l\,,
\qquad
{\cal G}_n\equiv\sum_{l=1}^\nu
P_{nl}(\gamma{\bf a}_l^\vphhh)\,,
$$
where

\begin{equation}\label{project}
P_{nl}
=\sum_{j=0}^{n-1}(-)^j\!\!\!\!\sum_{l_1\ldots l_{j+1}=1}^\nu
\delta_{nl_{j+1}}\delta_{l_1l}
\prod_{i=1}^j
\frac{\varepsilon^{l_i l_{i+1}}}{\lambda_{l_i n}}
{\bf a}_{l_{i+1}l_i}^\vphhh \,,
\end{equation}
$$
\lambda_{ln}^\vphhh
\equiv m_l^\vphhh - m_n^\vphhh +n-l +1\,.
$$
In (\ref{project}) the ${\bf a}_{l_{i+1}l_i}$ are ordered as
follows:
${\bf a}_{l_{j+1}l_j} ^\vphhh\ldots {\bf a}_{l_2l_1}^\vphhh\,.$
As an illustration of (\ref{gaugetr3}) we write down
$\delta_{(n)}|\Psi\rangle$ for $n=1,2,3$ assuming that such an
$n$ belongs to ${\sf S}({\bf m}^\prime)$:

\begin{eqnarray*}
&&\delta_{(1)}|\Psi\rangle
=D_1^\vphhh|\Lambda_1\rangle+(\gamma{\bf a}_1)|S_1\rangle\,,
\\
&&
\delta_{(2)}|\Psi\rangle
=D_2^\vphhh|\Lambda_2\rangle+(\gamma{\bf a}_2)|S_2\rangle
-\frac{{\bf a}_{21}}{\lambda_{12}}
(D_1^\vphhh|\Lambda_2\rangle+(\gamma{\bf a}_1)|S_2\rangle)
\,,
\\
&&
\delta_{(3)}|\Psi\rangle
=D_3^\vphhh|\Lambda_3\rangle+(\gamma{\bf a}_3)|S_3\rangle
-\frac{{\bf a}_{31}}{\lambda_{13}}
(D_1^\vphhh|\Lambda_3\rangle+(\gamma{\bf a}_1)|S_3\rangle)
-\frac{{\bf a}_{32}}{\lambda_{23}}
(D_2^\vphhh|\Lambda_3\rangle+(\gamma{\bf a}_2)|S_3\rangle)
\\
&&\hspace{1.5cm}
+\frac{{\bf a}_{32}
{\bf a}_{21}}{\lambda_{13}\lambda_{23}}
(D_1^\vphhh|\Lambda_3\rangle+(\gamma{\bf a}_1)|S_3\rangle)
\,.
\end{eqnarray*}

(iii) from the invariance requirement of eq.(\ref{gtrans})
(i.e.  $\bar{\gamma}_i\delta_{(n)}|\Psi\rangle=0$) we get the
equation for $|\Lambda_n\rangle$ and $|S_n\rangle$

\begin{equation}\label{gpeqmot1}
(\nline -(m_n^\prime-n)\yline)|\Lambda_n\rangle
+(d+2(m_n^\prime-n))|S_n\rangle=0
\end{equation}
and from the invariance requirement of eq.(\ref{diver})
(i.e. $\bar{\nabla}_i\delta_{(n)}|\Psi\rangle=0$) we get the
equation

\begin{equation}\label{gpeqmot11}
(\nabla^2-(m_n^\prime-n)(m_n^\prime-n-1+d))|\Lambda_n\rangle
+(\nline+(m_n^\prime-n+d)\yline)|S_n\rangle=0\,.
\end{equation}
Finally from the invariance requirement of eq.(\ref{eqmot2})
(i.e.  $(\nline^2-\ldots)\delta_{(n)}|\Psi\rangle=0$)
and from Eqs.(\ref{gpeqmot1}), (\ref{gpeqmot11}) we
derive the equation for $E_0$

\begin{equation}\label{quadeq}
(E_0+\frac{1}{2})(E_0+\frac{1}{2}-d)
=(m_n^\prime-n-1)(m_n^\prime-n-1+d)\,,
\quad n\in {\sf S({\bf m}^\prime)}\,,
\end{equation}
whose solutions read as

\begin{equation}\label{posenval}
E_{0(n)}^{(1)}=m_n-n-2+d\,,
\qquad
E_{0(n)}^{(2)}=n+1-m_n\,,
\qquad
n\in {\sf S}({\bf m}^\prime)\,.
\end{equation}
As seen from (\ref{posenval})
there exists an arbitrariness of $E_0$
parametrized by subscript $n$ which labels gauge
transformations and by superscripts $(1),(2)$ which label two
solutions of quadratic equation (\ref{quadeq}).
Because the values of $E_0$
have been derived by exploiting gauge invariance
we can conclude that the gauge invariance by itself does not
uniquely determine  the physical relevant value of $E_0$ (in this
regard the situation is similar to the $d=4$ case, see
\cite{F1}).  To choose physical relevant value of $E_0$ we
exploit the unitarity condition, that is: 1) hermiticity
$({\rm i}J^{AB})^\dagger={\rm i}J^{AB}$; 2) the positive norm
requirement.  The resulting procedure is the same that has been
previously encountered in the bosonic case \cite{Metsit2}.
Therefore we do not give deal of technical details and formulate
the result.

Given ${\sf YT}({\bf m}^\prime)$ let $k$, $k=1\ldots\nu$,
indicates maximal number of upper rows which have the same
number of boxes. We call such Young tableaux the level-$k$
${\sf YT}({\bf m}^\prime)$.  For the case of level-$k$ Young
tableaux the inequalities (\ref{mineq1}) can be rewritten as

\begin{equation}\label{mineq2}
m^\vphhh_1=\ldots =m_k^\vphhh>m_{k+1}^\vphhh \ge
m^\vphhh_{k+2}\ge \ldots \ge m_\nu^\vphhh\geq 1/2\,.
\end{equation}
Then making use of unitarity condition one proves
\cite{Metsit2} that for the level-$k$ Young tableaux the
$E_0$ should satisfy the inequality

\begin{equation}\label{enbound}
E_0\ge m_k^\vphhh-k-2+d\,.
\end{equation}
Comparing (\ref{posenval}) with (\ref{enbound})  we
conclude that only $E_{0(n=k)}^{(1)}$ satisfies the unitarity
condition.  Thus anti-de Sitter fermionic massless particles
described by level-$k$ ${\sf YT}({\bf m}^\prime)$ takes lowest
value of energy equal to

\begin{equation}\label{envaltr}
E_0=m_k-k-2+d\,.
\end{equation}

Note that it is gauge transformation with $n=k$
(\ref{gaugetr3}) that leads to relevant $E_0$, i.e.
given level-$k$ ${\sf YT}({\bf m}^\prime)$ only the gauge
transformation $\delta_{(k)}$ respects the unitarity.
Therefore only the $\delta_{(k)}$ will be used in what follows.
From now on we use letter $k$ to indicate level
of ${\sf YT}({\bf m}^\prime)$.  Our result for $E_0$
(\ref{envaltr}) cannot be extended to cover the case
${\bf m}^\prime=0$ because the $E_0$ obtained are relevant only for
gauge fields.  This case should be considered in its own right
and the relevant value $E_0(\hbox{for } {\bf
m}^\prime=0)=(d-1)/2$ can be obtained by using requirement of
conformal invariance (see \cite{Metsit3}).  The case ${\bf
m}^\prime=0$, $d=4$ has been investigated in \cite{DKS}.

Up to now we analysed second-order equations of motion for $\Psi$
and gauge invariance of these equations.  Now we would like to
derive first-order equations of motion and corresponding gauge
transformations.  To do that we use the relations

\begin{equation}\label{dirop}
\nline^2=\kappa^2-d\kappa\,,
\qquad
\kappa\equiv \sigma^{AB}L^{AB}\,,
\qquad
\kappa=\yline \nline\,,
\end{equation}
and rewrite equation of motions (\ref{eqmot2}) as follows:

$$
(\kappa-E_0-\frac{1}{2})(\kappa+E_0+\frac{1}{2}-d)|\Psi\rangle
=0\,.
$$
Thus we could use the following first-order equations
of motion:

\begin{equation}\label{eqmot3}
(\kappa+E_0+\frac{1}{2}-d)|\Psi\rangle
=0\,,
\end{equation}
\begin{equation}\label{eqmot4}
(\kappa-E_0-\frac{1}{2})|\tilde{\Psi}\rangle
=0\,.
\end{equation}
Due to relation

$$
(\kappa+E_0+\frac{1}{2}-d)|\Psi\rangle
=\yline(\kappa-E_0
-\frac{1}{2})\yline|{\Psi}\rangle
$$
it is clear that $|\Psi\rangle$ and $|\tilde{\Psi}\rangle$ are
related by
$|\Psi\rangle=\yline|\tilde{\Psi}\rangle$,
i.e. equations (\ref{eqmot3}) and (\ref{eqmot4}) are
equivalent. We will use the equation of motion given by
(\ref{eqmot3}). Now we should verify  gauge invariance of
the first-order equation (\ref{eqmot3}) with respect to gauge
transformations (\ref{gaugetr3}). It turns out that the
invariance requirement of (\ref{eqmot3}) with respect to
(\ref{gaugetr3}) leads to

\begin{equation}\label{ssol}
|S_k\rangle=0\,.
\end{equation}
Thus because of (\ref{mineq2}) and (\ref{ssol}) the final form
of gauge transformations is

\begin{equation}\label{gaugetr4}
\delta_{(k)}|\Psi\rangle
=\sum_{j=0}^{k-1}(-)^j\!\!\!\!\sum_{l_1\ldots l_{j+1}=1}^\nu
\delta_{kl_{j+1}}
\prod_{i=1}^j
\frac{\varepsilon^{l_i l_{i+1}}
{\bf a}_{l_{i+1}l_i}^\vphhh}{k+1-l_i}
D_{l_1}|\Lambda_k\rangle\,.
\end{equation}
The  equations of motion for $\Lambda$ can be obtained
from (\ref{gpeqmot1}) and (\ref{ssol})

\begin{equation}\label{gpeqmot2}
(\nline- (m_k-\frac{1}{2}-k)\yline)|\Lambda_k\rangle=0
\end{equation}
which can also be rewritten in terms of $\kappa$ and
$E_0^\Lambda$

\begin{equation}\label{gpeqmot3}
(\kappa +E_0^\Lambda+\frac{1}{2}-d)|\Lambda_k\rangle=0\,,
\end{equation}
where we introduce lowest energy value for $|\Lambda_k\rangle$:
$E_0^\Lambda=E_0+1$

\begin{equation}\label{gpenval}
E_0^\Lambda=m_k-k-1+d\,.
\end{equation}
With the values for $E_0^\Lambda$ at hand we are ready to
provide an answer to the question: do the gauge parameter
fields meet the masslessness criteria?
Because the inter-relation between of spin ${\bf m}$ and
energy value $E_0$ for massless field is given by
(\ref{envaltr}) we should express the $E_0^\Lambda$ in terms
of ${\bf m}^\Lambda$ and $k^\Lambda$, where $k^\Lambda$ is a
level of ${\sf YT}({\bf m}^{\prime\Lambda})$. Due to relations
$k^\Lambda=k-1$, $m^\Lambda_{k^\Lambda}=m_k-\delta_{k1}$ we
transform (\ref{gpenval}) to

$$
E_0^\Lambda=m_{k^\Lambda}^\Lambda-k^\Lambda-2+d+\delta_{k1}\,.
$$
Comparing this relation with (\ref{envaltr}) we conclude that
only for $k>1$ the gauge parameters are massless fields while
for $k=1$ they are massive fields.

Thus we have constructed equations of motion (\ref{eqmot3}) which
respect gauge transformations (\ref{gaugetr4}), where the
gauge parameter fields $\Lambda$ satisfy the constraints
(\ref{gpcon1}), (\ref{ytgpcon2}) and equations of motion
(\ref{gpeqmot3}).  The relevant values of $E_0$ and
$E_0^\Lambda$ are given by (\ref{envaltr}) and (\ref{gpenval}).
Note that the equations of motion are written in terms of the
operator $\kappa$ (see equation (\ref{dirop})) introduced in
\cite{Dir} while constructing the equation of motion for the
field associated with the representation $D(E_0,1/2)$. The
$\kappa$ is expressible in terms of the orbital momentum $L^{AB}$
(see equation (\ref{dirop})).  Now we would like to rewrite our
equations of motion in terms of complete angular momentum
$J^{AB}$. We believe that such a formulation  will form a good
basis for establishing an action leading to the equations of
motion under consideration.  To this end let us first multiply
equations (\ref{eqmot3}) and (\ref{gpeqmot3}) by $\yline$. Then
making use of equalities

\begin{eqnarray}
\label{jeq1}
&&
\gamma^Ay^BJ^{AB}=\nline+\sum_{l=1}^\nu(
\gamma_l^\vphhh\bar{y}_l^\vphhh
-y_l^\vphhh\bar{\gamma}_l^\vphhh)+\frac{1}{2}d\yline\,,
\\
\label{jeq2}
&&
a_n^Ay^BJ^{AB}=D_n+\frac{1}{2}a_n^A\theta^{AB}\gamma^B\yline\,,
\end{eqnarray}
and the constraints (\ref{ytrans}), (\ref{gtrans}),
(\ref{gpcon1}) we get the desired form for equations of motion:

\begin{equation}\label{eqmot5}
\Bigl(\gamma^Ay^BJ^{AB}-(E_0-\frac{d-1}{2})\yline
\Bigr)|\Psi\rangle=0\,,
\end{equation}
\begin{equation}\label{gpeqmot5}
\Bigl(\gamma^Ay^BJ^{AB}
-(E_0^\Lambda-\frac{d-1}{2})\yline\Bigr)|\Lambda_k\rangle
=0\,,
\end{equation}
where $E_0$ and $E_0^\Lambda$  are given by
(\ref{envaltr}) and (\ref{gpenval}).
Because the term $D_{l_1}|\Lambda_k\rangle$ from gauge
transformations (\ref{gaugetr4}) can be expressed as

\begin{equation}\label{gaugetr7}
D_l|\Lambda_k\rangle
=(a_l^A-\frac{a_l^C\theta^{CD}\gamma^D}{2E_0+3-d}\gamma^A)
y^BJ^{AB}|\Lambda_k\rangle\,,
\end{equation}
it is seen that the gauge transformations (\ref{gaugetr4}) are
also expressible in terms of $J^{AB}$.
Note that in deriving of
(\ref{gpeqmot5}) and (\ref{gaugetr7}) it is necessary to use the
equations of motion for $|\Lambda_k\rangle$ (\ref{gpeqmot3}) and
the relevant values of $E_0$ and $E_0^\Lambda$ given by
(\ref{envaltr}) and (\ref{gpenval}).

Now we would like to transform our results to intrinsic
coordinates, in terms of covariant derivatives and vierbein
fields. Let $x^\mu$, $\mu=0,1,\ldots, d-1$ be the intrinsic
coordinates for anti-de Sitter spacetime and let $y^A(x)$ be
imbedding map, where $y^A(x)$ satisfy (\ref{hyperbol}).  The
relationship between $d+1$-dimensional tensor-spinor field
$\Psi^{A\ldots}$ and the usual Rarita-Schwinger tensor-spinor
field $\psi_{\mu\ldots}$ is given by (see \cite{FF})

$$
\psi_\mu\ldots(x)=M^{-1}(x)y^A_\mu\ldots \Psi^{A\ldots}(y)\,,
$$
where $y_\mu^A\equiv \partial_\mu y^A$ and $M$ is an
$2^{d/2}\times 2^{d/2}$ matrix defined by

\begin{equation}\label{mmatrix}
M^{-1}\partial_\mu M
=\frac{1}{2}\omega_\mu^{ab}\sigma_{ab}
+\frac{1}{2}e_\mu^a\gamma_a\,,
\qquad
M^{-1}(\yline\partial_\mu \yline) M
=e_\mu^a \gamma_a\,,
\end{equation}
$\sigma^{ab}=(\gamma^a\gamma^b-\gamma^b\gamma^a)/4$,
where $a,b=0,1,\ldots,d-1$ are the tangent space indices and
$\gamma^a$ are the gamma matrices
$\{\gamma^a,\gamma^b\}=2\eta^{ab}$. The   $e_\mu^a$ and
$\omega_\mu^{ab}$ are the vierbein and Lorentz connection of
anti-de Sitter spacetime.  Concrete representation for matrix
$M$ may be found in (\cite{FF}). As usual it is convenient to
introduce generating function for Rarita-Schwinger tensor-spinor
field

\begin{equation}\label{defpsi}
|\psi\rangle
=M^{-1}\prod_{l=1}^\nu P_l^{m_l^\prime}|\Psi\rangle\,,
\qquad
P_l\equiv a_l^be_b^\mu y_\mu^A \bar{a}_l^A\,,
\end{equation}
where $a_l^b$ are new creation operator
$[\bar{a}_i^a,a_j^b]=\delta_{ij}\eta^{ab}$,
$\eta^{ab}=(-,+,\ldots,+)$.  Note that the $|\psi\rangle$ is
actually a generating function for tangent space
Rarita-Schwinger field:
$\psi_{a\ldots}=e^\mu_a\ldots\psi_{\mu\ldots}$. Now making use of
(\ref{mmatrix}) the equations of motion (\ref{eqmot3}) can be
rewritten in terms of $|\psi\rangle$ as follows

\begin{equation}\label{eqmot6}
(\gamma^a e_a^\mu D_{\mu L}+E_0+\frac{1-d}{2})|\psi\rangle
=0\,,
\end{equation}
$$
D_{\mu L}\equiv \partial_\mu+\frac{1}{2}\omega_\mu^{ab}M_{ab}\,,
\qquad
M^{ab}\equiv\sum_{l=1}^\nu (a_l^a\bar{a}_l^b-a_l^b\bar{a}_l^a)
+\sigma^{ab}\,.
$$
In order to transform the gauge transformation to intrinsic
coordinates we introduce the gauge parameter $|\lambda_k\rangle$
by analogy with (\ref{defpsi})

\begin{equation}\label{deflam}
|\lambda_k\rangle=M^{-1}\prod_{l=1}^\nu
P_l^{m_{l(k)}^\prime}|\Lambda_k\rangle\,,
\end{equation}
and now the gauge transformation (\ref{gaugetr4}) is rewritten
as

\begin{equation}\label{gaugetr5}
\delta_{(k)}|\psi\rangle
=\sum_{j=0}^{k-1}(-)^j\!\!\!\!\sum_{l_1\ldots l_{j+1}=1}^\nu
\delta_{kl_{j+1}}
\prod_{i=1}^j
\frac{\varepsilon^{l_i l_{i+1}}
(a_{l_{i+1}}^\vphhh \bar{a}_{l_i}^\vphhh)}{k+1-l_i}
a^b_{l_1}
(e_b^\mu D_{\mu L}+\frac{1}{2}\gamma_b)
|\lambda_k\rangle\,,
\end{equation}
where $(a_ia_j)\equiv \eta_{ab}a^aa^b$.
Note that the $|\lambda_n\rangle$ satisfies the equation of motion
which is obtainable from (\ref{eqmot6}) by making there the
substitutions $|\psi\rangle\rightarrow |\lambda_k\rangle$ and
$E_0\rightarrow E_0^\Lambda$.
The constraints (\ref{diver}), (\ref{gtrans})
take the form

\begin{equation}\label{finfcon}
\bar{a}_n^b(e_b^\mu D_{\mu L}+\frac{1}{2}\gamma_b)|\psi\rangle
=0\,,
\qquad
\bar{a}_n^b\gamma_b|\psi\rangle=0\,,
\end{equation}
$n=1,\ldots,\nu$, and similar constraints  for
$|\lambda_k\rangle$ are obtainable from (\ref{finfcon}) by
making there the substitution $|\psi\rangle\rightarrow
|\lambda_k\rangle$.  The constraints (\ref{ytfcon2})
transform to

\begin{equation}\label{ytfcon3}
((a_i \bar{a}_i)-m_i^\prime)|\psi\rangle=0\,,
\qquad
(\bar{a}_i \bar{a}_i)|\psi\rangle=0\,,
\qquad
\varepsilon^{ij}(a_i \bar{a}_j)|\psi\rangle=0\,,
\end{equation}
$i,j=1,\ldots,\nu$, while similar constraints for
$|\lambda_k\rangle$ are obtainable from (\ref{ytfcon3}) by
making there the substitutions $|\psi\rangle\rightarrow
|\lambda_k\rangle$ and $m_i^\prime\rightarrow m_{i(k)}^\prime$.

In conclusion let us summarize the results of this letter.
For massless fermionic field $|\Psi\rangle$ of arbitrary spin
labeled by ${\bf m}$
we have constructed: (i) free wave equations of motion
in $SO(d-1,2)$ covariant form (\ref{eqmot3})
as well as in terms of intrinsic coordinates (\ref{eqmot6});
(ii) corresponding gauge transformations (\ref{gaugetr4}),
(\ref{gaugetr5}), subsidiary conditions
(\ref{diver})-(\ref{gtrans}), (\ref{ytfcon2}),
(\ref{ytgpcon2}), (\ref{finfcon}),
(\ref{ytfcon3}) and equations of motion for gauge parameter
(\ref{gpeqmot3}) field written also in both forms;
(iii) the new representation for equations of
motion (\ref{eqmot5}),(\ref{gpeqmot5}) and gauge transformations
(see (\ref{gaugetr4}),(\ref{gaugetr7})) in terms of the
generators of the anti-de Sitter group $SO(d-1,2)$;
(iv) lowest energy values for massless fermionic field
(\ref{envaltr}) and for gauge parameter field (\ref{gpenval}).
Also we have demonstrated that the gauge parameter for massless
field associated with level $k=1$ Young tableaux is a massive
field while for $k>1$ the gauge parameter is a massless
field.

The author would like to thank Prof. B. de Wit for hospitality
at Institute for Theoretical Physics of Utrecht University where
a part of this work was carried out. This work was supported
in part by INTAS contracts  No. CT93-0023 and No.96-538, by the
Russian Foundation for Basic Research, Grant No.96-01-01144,
and by the NATO Linkage, Grant No.931717.

\end{document}